\documentclass[aps,prl,twocolumn,floatfix]{revtex4}
\usepackage{graphicx}
\usepackage{amsmath}
\usepackage{amssymb}

\newcommand{\rs}{\mathrm{sat}}
\newcommand{\rms}{\mathrm{rms}}

\begin{document}
\title{Continuous and Discontinuous Routes to Chaos with Diverging Transport Times}
\author{Anne K\'etri P. da Fonseca$^{1,2}$, Marcelo de Almeida Presotto$^2$, Diego F. M. Oliveira$^1$, Edson D.\ Leonel$^2$}

\affiliation{$^1$School of Electrical Engineering and Computer Science, University of North Dakota, USA\\
$^2$Departamento de F\'isica, Unesp - Universidade Estadual Paulista - 
Av.24A. 1515, 13506-900, Rio Claro, SP, Brazil}

\date{\today}

\begin{abstract}
The onset of chaos and the exploration of phase space need not occur on the same timescale. We quantify this distinction in a stadium-like billiard map whose rectangular integrable limit admits two contrasting responses to opposite boundary deformations. For dispersing boundaries, the local exponential instability time at fixed periodic orbits grows as the inverse square root of the deformation, whereas numerical angular spreading approaches a finite plateau on a timescale growing approximately as its inverse square. Their ratio is predicted to diverge as the deformation to the power $-3/2$ if the angular diffusivity vanishes quadratically. An exactly solvable confined random walk isolates the kinetic origin of the slowing down. A scaling criterion separates the width sampled by the dynamics from its exploration rate, clarifying when a stationary discontinuity can coexist with a divergent transport time. These results connect the dynamical onset of chaos to mixed-order phenomenology and distinguish microscopic instability from macroscopic transport.
\end{abstract}
\maketitle

The distinction between continuous and discontinuous phase transitions is commonly expressed in terms of an order parameter \cite{landaunature,goldenfeld,sethna2021statistical}. Yet a finite jump need not exclude critical singularities: mixed-order transitions in long-range Ising models and hybrid transitions in network percolation combine discontinuous order parameters with divergent correlation scales or singular fluctuations \cite{thouless1969,bar2014mixed,dorogovtsev2006kcore,goltsev2006kcore}. These examples motivate a dynamical question: can distinct stationary responses, one continuous and one discontinuous, emerge from opposite perturbations of the same integrable system? And can the onset of local instability be separated quantitatively from the time required to explore phase space?

Near integrability, resonances, islands and invariant curves restrict transport in Hamiltonian systems \cite{MEISS,leonel2015dynamical,chernov2006chaotic}. The accessible angular or action width can shrink continuously as the perturbation vanishes, as observed in scaling studies of maps and billiards \cite{leonel2020characterization,daFonseca2026}. Alternatively, the width can remain finite even while the transport coefficient tends to zero. The two stationary responses have previously been identified in this stadium-like map \cite{daFonseca2026}. Here we ask what sets the separation between the onset of exponential instability and the much slower angular exploration, and which features of the stationary response are attributable to geometry rather than to the rate of transport. We use a bounded stochastic walk as an analytically tractable reference, not as a realization of deterministic chaos. The numerical results for the two models were also reported in a related preprint \cite{daFonsecaLong2026}; the present analysis focuses on the separation of instability and exploration scales and on the discrete spectral and correlation-dependent transport results.

\begin{figure}[t]
 \includegraphics[width=\columnwidth]{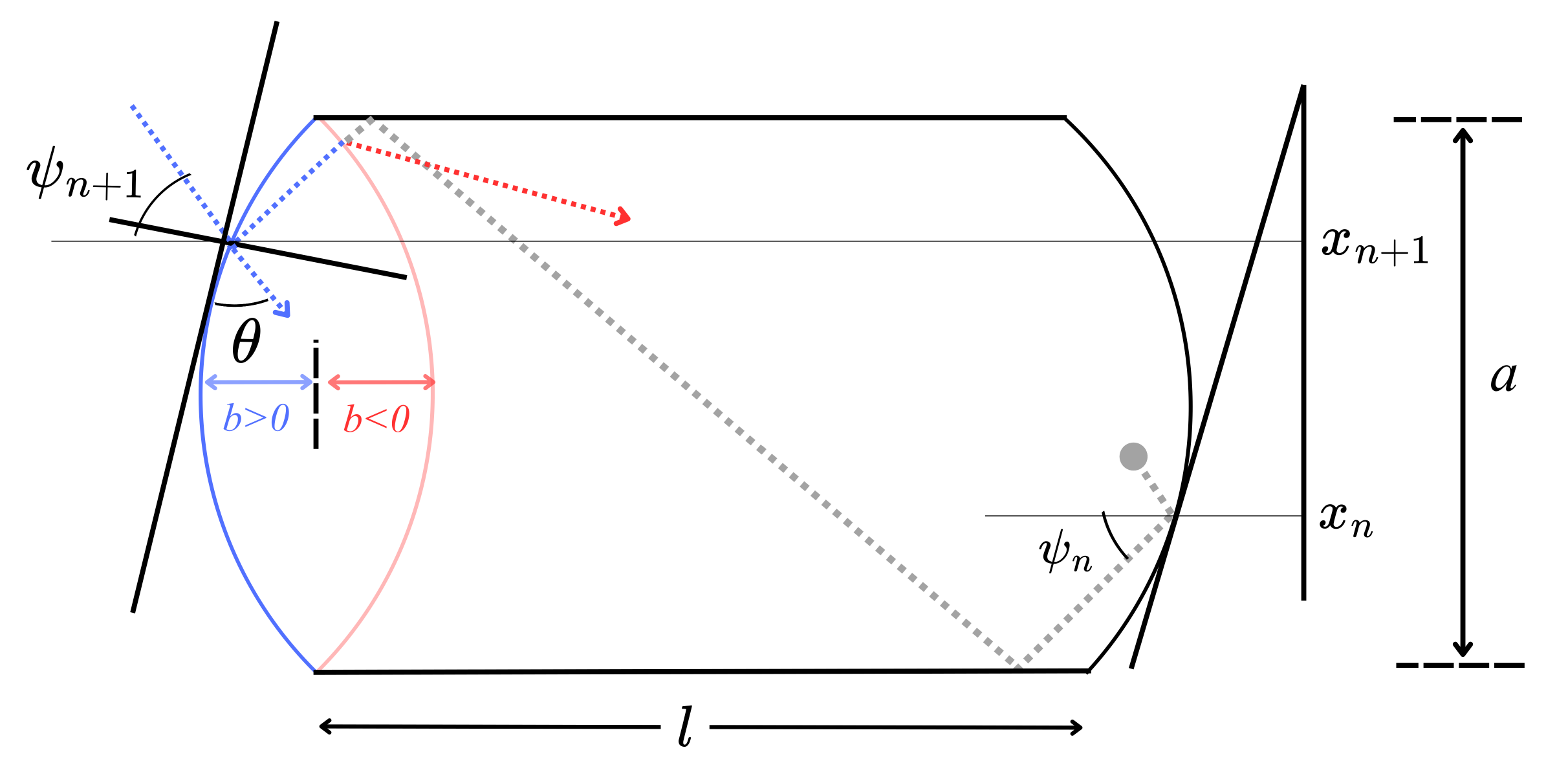}
 \caption{Stadium-like geometry and unfolded collision variables. The sign of $b$ distinguishes focusing ($b>0$) and dispersing ($b<0$) end caps; $b=0$ gives the rectangle. Here $a$ is the transverse height and $l$ the axial separation.}
 \label{fig:geometry}
\end{figure}

Consider a stadium-like billiard with parabolic end caps separated by a length $l$, of transverse height $a$ [Fig.~\ref{fig:geometry}]. The signed deformation $b$ is positive for focusing and negative for dispersing boundaries in our convention. At $b=0$ the caps are flat and the rectangle is integrable. Following the unfolded, small-curvature approximation of Refs.~\cite{loskutov2002,livorati2011family}, valid for $|b|\ll l$ and small slopes $|b|/a$, the map between successive end-cap encounters is
\begin{equation}
\begin{split}
 \xi_{n+1}&=\left(\xi_n+\frac{l}{a}\tan\psi_n\right)\bmod 1,\\
 \psi_{n+1}&=\psi_n-\frac{8b}{a}(2\xi_{n+1}-1),
\end{split}\label{eq:billiard}
\end{equation}
where $n$ counts map iterations (collision counts, not physical flight time), $\xi_n=x_n/a\in[0,1)$ is the normalized transverse coordinate and $\psi_n$ is the signed flight angle relative to the axial direction. A dynamical billiard describes a point particle moving freely between specular reflections at a boundary. The angular update is local, away from grazing and the cut at $\pm\pi/2$. The angular continuation affects both the invariant measure and the measured roughness (Supplemental Material \cite{supplement}). We study this reduced map, not the exact specular billiard.

\begin{figure}[t]
 \includegraphics[width=\columnwidth]{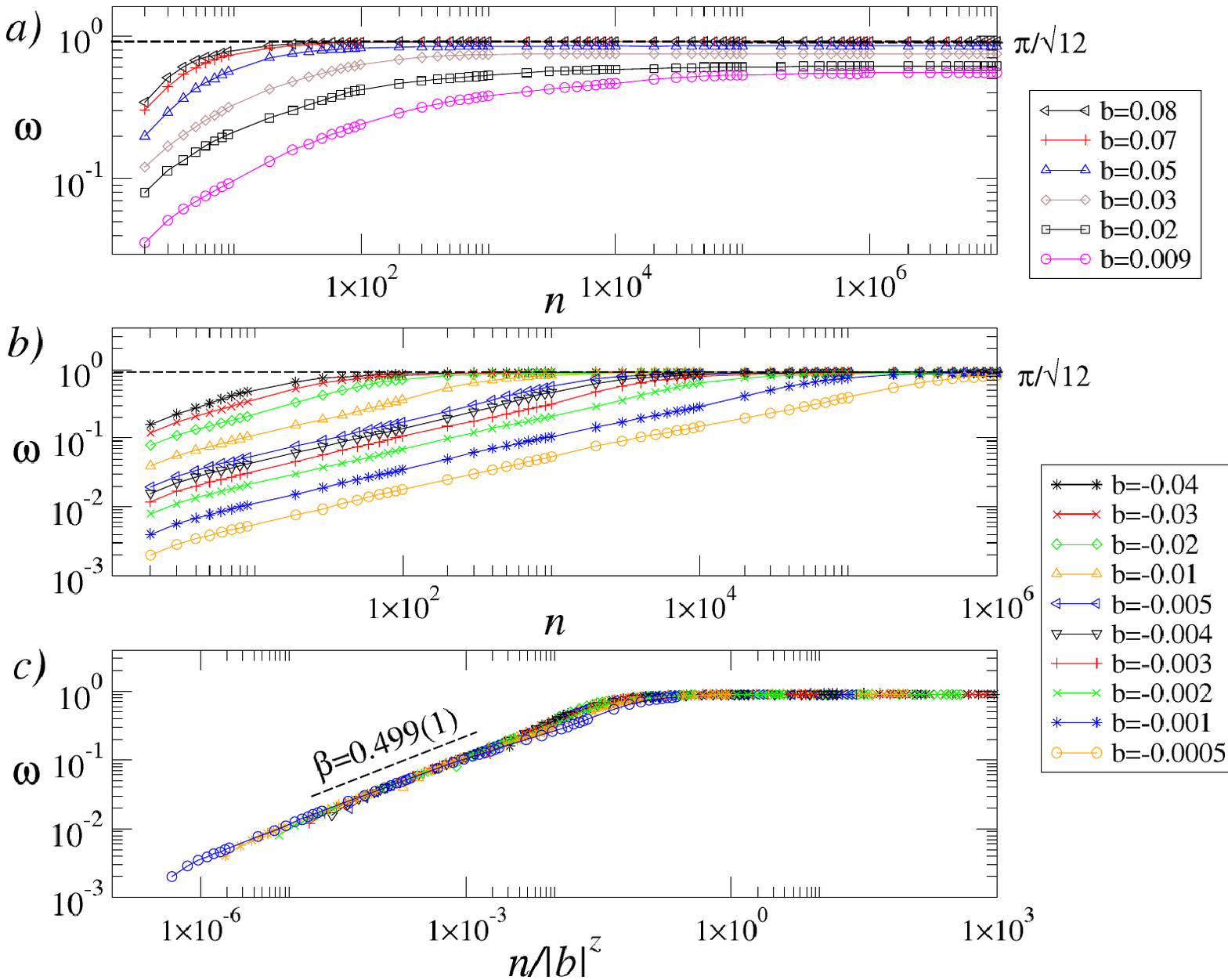}
 \caption{Angular roughness from Eq.~(\ref{eq:omega}). (a) Focusing branch ($b>0$): the stationary width varies with $b$. (b) Dispersing branch ($b<0$): approximately common plateau near the uniform-angle reference $\pi/\sqrt{12}$. (c) Dispersing data collapsed using $n/|b|^z$ with $z\simeq-2$; the displayed early-time slope is $0.499(1)$. The numerical series is that reported in Ref.~\cite{daFonsecaLong2026}; the collapse uses the rounded exponent $z=-2$.}
 \label{fig:billiard}
\end{figure}

For $M$ trajectories, indexed by $j$, define the finite-time roughness
\begin{equation}
 \begin{split}
 \omega(n,b)&=\frac1M\sum_{j=1}^M
 \sqrt{\overline{\psi_j^2}-\overline{\psi_j}^{\,2}},\\
 \overline{\psi_j^k}&=\frac1n\sum_{i=1}^n\psi_{i,j}^k .
 \end{split}
 \label{eq:omega}
\end{equation}
Thus $\omega$ averages the temporal standard deviation of each orbit, not the ensemble variance at one iteration. We define $\omega_\rs(b)=\lim_{n\to\infty}\omega(n,b)$ for the specified ensemble. At $b=0$ each orbit has constant $\psi$, hence $\omega(n,0)=0$ even for an ensemble with different initial angles. For the data of Fig.~\ref{fig:billiard}, $a=0.5$, $l=1$, and $M=10^3$ trajectories are sampled with initial $\xi$ distributed uniformly and with initial angles in the component under investigation. The stationary roughness serves as an order-parameter-like measure of angular exploration.

The integrable limit is particularly instructive because the entire family of invariant lines $\psi=\mathrm{const.}$ is present in the same phase space. On each line the transverse coordinate advances by the rotation number $(l/a)\tan\psi$ modulo unity: rational rotation numbers give periodic motion, whereas irrational ones can sample the transverse coordinate densely without any angular transport. Thus even extensive exploration of $\xi$ at $b=0$ is compatible with zero angular roughness. The perturbation couples the transverse coordinate back to the angle and removes the conservation of $\psi$ for generic orbits. The question is then whether the typical angular dispersion decreases toward zero or remains finite even as individual angular kicks vanish. This distinction concerns the angular measure sampled by the trajectories and need not imply that the connected chaotic region itself shrinks.

On the local branches of map (\ref{eq:billiard}), the period-one fixed points with zero angular increment are $\xi^*=1/2$, $\psi_m^*=\arctan(ma/l)$ with integer $m$. The Jacobian has unit determinant and trace $2-16bl/(a^2\cos^2\psi_m^*)$. For sufficiently small positive $b$ at fixed $m$, these points are elliptic; for negative $b$ they are hyperbolic. The central focusing fixed point changes linear stability at $b_c=a^2/(4l)$; it does not disappear. In the dispersing branch the local instability exponent per iteration at fixed $m$ is (Supplemental Material \cite{supplement})
\begin{equation}
 \begin{split}
 \gamma_m&=\operatorname{arcosh}\!\left(1+\frac{8|b|l}{a^2\cos^2\psi_m^*}\right)\\
 &\sim\frac{4\sqrt{l|b|}}{a|\cos\psi_m^*|}.
 \end{split}\label{eq:local}
\end{equation}
The associated e-folding time $n_{\rm inst}=1/\gamma_m$ therefore diverges as $|b|^{-1/2}$. On the focusing side regular islands and chaotic regions coexist; on the dispersing side the phase portraits show extensive angular exploration in the sampled region \cite{daFonseca2026}. The asymmetry is geometrical and is already visible in the stability of the same family of fixed points.

\begin{figure}[t]
 \includegraphics[width=\columnwidth]{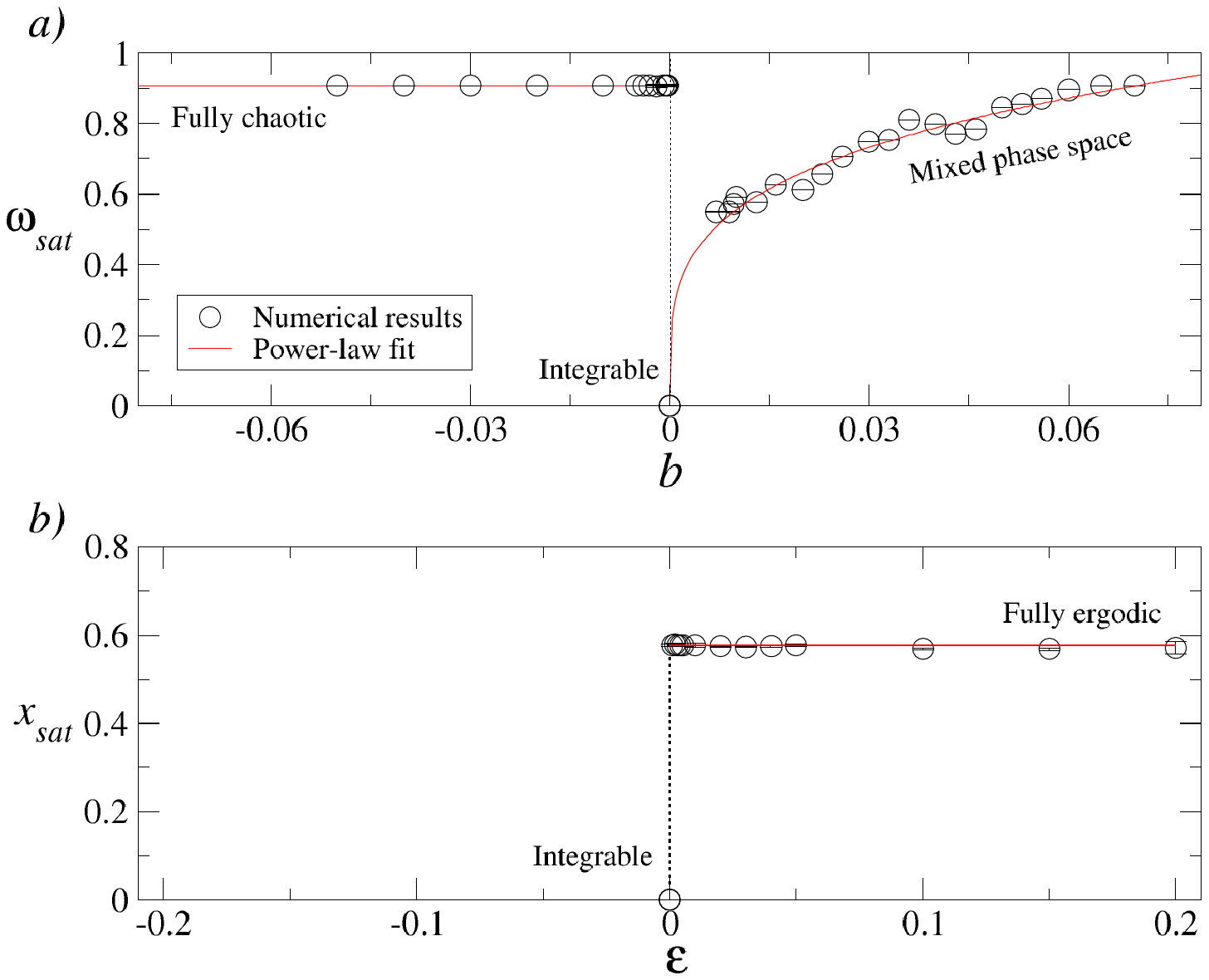}
 \caption{Stationary spreading. (a) Billiard: continuous focusing branch and finite dispersing plateau. (b) Confined walk: $x_\rs(0)=0$ for $x_0=0$ and $x_\rs=L/\sqrt3$ for nonzero steps with specular reflection.  The original plot labels denote frozen motion in (b) at zero steps and observed full-range spreading on the plateaus, not a proof of ergodicity. Symbols show numerical results previously reported in Ref.~\cite{daFonsecaLong2026}; lines guide the eye or show scaling fits.}
 \label{fig:order}
\end{figure}

The focusing and dispersing branches show different stationary responses (Figs.~\ref{fig:billiard} and \ref{fig:order}). On the focusing side, a decreasing stationary roughness was reported in Ref.~\cite{daFonseca2026}; the intermediate scaling regime gave a stationary-width exponent $0.28(2)$, while trapping near regular islands (stickiness) complicates extrapolation to smaller deformations. A small roughness characterizes the angular distribution sampled by a trajectory, rather than necessarily the geometric support of the entire connected chaotic component. On the dispersing side, the observed plateau is close to $\pi/\sqrt{12}$, the standard deviation of a uniform distribution on the principal angular interval. This value describes the reported reduced-map statistics. The numerical results motivate the asymptotic interpretation
\begin{equation}
 \lim_{b\to0^+}\omega_\rs(b)=0,\qquad
 \lim_{b\to0^-}\omega_\rs(b)\simeq\frac{\pi}{\sqrt{12}}\ne\omega_\rs(0).
 \label{eq:jumpB}
\end{equation}
The limits in Eq.~(\ref{eq:jumpB}) are extrapolations from the observed scaling and plateaus, not proofs of the deterministic invariant measures. The zero-deformation value is exact.

The comparison in Fig.~\ref{fig:order}(a) is especially useful because the geometric parameter changes sign while the integrable reference and the observable remain the same. In Fig.~\ref{fig:billiard}(a) the saturation levels separate as the deformation is reduced; in panel (b) the dominant effect is instead a horizontal displacement of the crossover toward larger $n$. The dispersing branch in panel (c) consequently admits a collapse by rescaling the iteration axis alone, to leading order. These are different signatures of integrability breaking: a change in the angular dispersion sampled on one side, and predominantly a change in the time needed to explore a finite range on the other. In a mixed phase space, the measured stationary roughness also depends on which invariant component is sampled; here, ensembles are initialized within the chaotic sea.

Before saturation, the dispersing roughness grows approximately as the square root of the iteration number, consistent with normal diffusion. Define the exponents by $\omega\propto n^\beta$ in the growth regime, $\omega_\rs\propto |b|^\alpha$ on the stationary branch and $n_x\propto |b|^z$, where $n_x$ is the crossover iteration between the growth and saturation regimes. Matching the two asymptotes supplies a natural theoretical definition of the crossover (other fixed-fraction definitions share its exponent under single-parameter scaling). The fits annotated in the original numerical panels give $\beta=0.499(1)$, $z=-2.05(1)$ and $\alpha=0.00008(2)$ (Supplemental Material \cite{supplement}). The values are numerically close to the ideal exponents $(\alpha,\beta,z)=(0,1/2,-2)$, but the stationary-width and crossover fits differ from their ideal values by more than their quoted fit uncertainties.  These slight departures reflect preasymptotic finite-$\vert{}b\vert{}$ and finite-time effects beyond statistical regression errors.

Away from the angular cut, the dispersing kick is $\Delta\psi_n=8|b|u_n/a$, with $u_n=2\xi_{n+1}-1$. If the fast variable is approximately uniform, $\langle u\rangle\simeq0$ and $\langle u^2\rangle\simeq1/3$, where brackets denote an average over the relevant orbit ensemble. The effective angular diffusivity $D_\psi$ is the coefficient of the linear growth of the centered, unconfined angular-increment variance, i.e., $\operatorname{Var}(\sum_{i=0}^{n-1}\Delta\psi_i)\simeq2D_\psi n$ in the preconfinement diffusive regime. Under a stationary fast-process approximation with summable correlations, it obeys the Green--Kubo estimate
\begin{equation}
 D_\psi=\frac{32b^2}{a^2}\left[C_b(0)+2\sum_{k\ge1}C_b(k)\right],
 \label{eq:Dpsi}
\end{equation}
with $C_b(k)=\langle(u_i-\langle u\rangle)(u_{i+k}-\langle u\rangle)\rangle$. Independent uniform kicks give $D_\psi\simeq32b^2/(3a^2)$. Quadratic scaling requires the bracket to tend to a finite positive limit. The observed collapse is consistent with this coarse-grained picture over the fitted range.

For a finite, perturbation-independent characteristic angular width, the diffusion approximation gives $n_x\propto D_\psi^{-1}\propto |b|^{-2}$ if the correlation factor has a finite positive limit. Combining this with Eq.~(\ref{eq:local}) gives the distinct-scale prediction
\begin{equation}
 \frac{n_x}{n_{\rm inst}}\propto |b|^{-3/2},\qquad b\to0^-,
 \label{eq:ratio}
\end{equation}
at any fixed $m$, under quadratic diffusivity scaling. Thus the local instability e-folding time cannot account for the much longer global exploration time.

There is a complementary physical interpretation of Eq.~(\ref{eq:ratio}). Hyperbolicity describes the amplification of an infinitesimal displacement in phase space; angular diffusion describes the accumulation of finite changes of the flight direction. These processes are not interchangeable. At a fixed hyperbolic periodic orbit, a decrease of $|b|$ weakens the instability, but the angular increments become small at a different rate. If the effective correlation factor in Eq.~(\ref{eq:Dpsi}) approaches a finite positive limit, the number of local instability times contained in one transport crossover grows without bound. The ratio is a comparison with the specified fixed-orbit scale. Moreover, in the nearly integrable limit the transverse dynamics may retain correlations associated with the unperturbed rotations. This is why the correlation factor, rather than the kick variance alone, is the relevant quantity for testing the transport prediction. The present collapse supplies evidence for normal transport over the explored range, leaving the limiting behavior of that factor as a distinct question.

To separate confinement and mobility from deterministic instability, consider independent walkers initially at $x_0=0$ on $[-L,L]$, with steps $x_{i+1}=R_L(x_i+\varepsilon Z_i)$, where $Z_i$ are independent uniform variables on $[-1,1]$, $\varepsilon\ge0$ is the maximum step length, and $R_L$ implements specular reflection at the endpoints. The exact results below refer to this reflection rule. Define
\begin{equation}
 x_\rms(n)=\left[\frac1{M}\sum_{j=1}^M\frac{1}{n}\sum_{i=1}^n x_{i,j}^2\right]^{1/2}.
 \label{eq:rw}
\end{equation}
Define $x_\rs(\varepsilon)=\lim_{n\to\infty}x_\rms(n,\varepsilon)$. For nonzero $\varepsilon$ the stationary density is uniform, giving $x_\rs=L/\sqrt3$; at $\varepsilon=0$ the localized preparation remains frozen [Fig. 3(b)]. In the diffusion limit $\varepsilon/L\to0$, $n\to\infty$ with dimensionless diffusive time $s = D n / L^2 = \varepsilon^2 n / (6L^2)$ fixed, the density obeys $\partial_n P=D\partial_x^2P$ with $D=\varepsilon^2/6$, Neumann conditions at $\pm L$, and $P(x,0)=\delta(x)$. The full Neumann gap is $\pi^2D/(4L^2)$; for the centered, even density and the second moment the first active decay rate is $\pi^2D/L^2$. The corresponding scale $\tau_{\rm even}=L^2/(\pi^2D)$ diverges as $\varepsilon^{-2}$. The exact discrete reflected-walk spectrum, derived in the Supplemental Material \cite{supplement}, yields the same limit without assuming a differential equation.

In the early regime, $\langle x_i^2\rangle=2Di$ before boundary effects, so the time average in Eq.~(\ref{eq:rw}) gives $x_\rms^2\simeq D(n+1)$, whereas $x_\rs^2=L^2/3$. This yields $\beta=1/2$, $\alpha=0$, $z=-2$ by matching. Here $\alpha$ describes the stationary width, $\beta$ the early growth in iteration number, and $z$ the divergence of the crossover iteration. The reported fits are $\beta=0.50025(5)$, $z=-2.004(5)$ and $\alpha=0.0028(7)$ (see the Supplemental Material \cite{supplement}). These fits are retained as annotated in the original panels; the slight departures of the fitted $\beta$ and $\alpha$ from the limiting predictions reflect finite-step ($\varepsilon/L > 0$) and finite-time ($D(n+1)$) preasymptotic corrections, together with finite-ensemble regression uncertainties. The exact stationary result and the asymptotic scaling refer to the specified specularly reflected process. This model illustrates loss of ergodicity at zero mobility.

The stochastic reference also illustrates the importance of specifying the preparation and the observable. For a walk started at a nonzero $x_0$, the uncentered second moment at zero mobility would give $x_\rms=|x_0|$, rather than zero; the discontinuity discussed here refers to the localized central preparation. By contrast, at $b=0$ the orbit-centered billiard roughness vanishes for every constant-angle trajectory, regardless of its initial angle. The agreement of the scaling exponents therefore does not imply that the two observables, their scaling functions, or their stationary measures are identical. It reflects the common structure of bounded spreading with a slow exploration rate.

Let $\lambda$ denote a dimensionless perturbation magnitude ($\varepsilon/L$ or $|b|/a$ here). Suppose that a spreading observable $W$ is controlled by a single characteristic sampled width $\ell(\lambda)$ and normal transport with diffusivity $D(\lambda)$, with a fixed rescaled geometry and preparation. If $\ell\sim\lambda^\kappa$ and $D\sim\lambda^p$, a single-scale diffusion description \cite{Family1985} gives
\begin{equation}
 W(n,\lambda)\sim \ell(\lambda)F\!\left(\frac{Dn}{\ell^2}\right),\quad
 F(s)\sim\begin{cases}c_0s^{1/2},&s\ll1,\\c_\infty,&s\gg1,\end{cases}
 \label{eq:general}
\end{equation}
where the constants and function depend on the observable, preparation and boundary conditions. Consequently,
\begin{equation}
 \alpha=\kappa,\quad\beta=\tfrac12,\quad z=2\kappa-p,
 \qquad \alpha-z\beta=p/2.
 \label{eq:relation}
\end{equation}
A stationary jump with a divergent crossover is obtained when $\kappa=0$, $p>0$, $c_\infty\ell(0^+)>0$ and $W(n,0)=0$. For $\kappa>0$ the stationary width instead vanishes continuously, even though the crossover can still diverge if $p>2\kappa$. The criterion separates the characteristic sampled width from the exploration rate. It is conditional on single-scale normal diffusion and is not a description of the sticky focusing regime.

In this description, the exponent $z$ is a composite transport exponent. A divergence of the crossover iteration alone cannot determine whether the stationary response is continuous: the accessible extent and diffusivity enter with different powers. Conversely, a nearly constant saturation value does not establish the rate at which the associated stationary measure is approached. Measuring these quantities separately is therefore important when comparing different systems. In the dispersing map, residual dynamical correlations can also contribute to the preasymptotic deviations noted above.

A further prediction of this scaling description is finite-observation rounding. For the discontinuous branch, the limits obey $\lim_{\lambda\to0^+}\lim_{n\to\infty} W(n,\lambda)>0$ but $\lim_{n\to\infty}\lim_{\lambda\to0^+}W(n,\lambda)=0$ for localized or orbit-centered preparations away from the angular cut. At finite $n$, the stationary jump is rounded across a perturbation scale $\lambda_n\sim n^{-1/p}$, hence $n^{-1/2}$ for quadratic diffusivity. This distinguishes a stationary jump from incomplete exploration. Time-averaged and instantaneous relaxation need not coincide (Supplemental Material \cite{supplement}).

The inferred stationary jump and divergent transport time resemble mixed-order phenomenology \cite{bar2014mixed,dorogovtsev2006kcore,goltsev2006kcore}, but its origin differs from diverging spatial correlations or collective percolation fluctuations. Our domain is finite and the spectral slowing of the stochastic reference follows from vanishing mobility. In the deterministic map, boundary geometry shapes the angular distribution sampled by the trajectories, and the correlations of the small kicks enter the transport scale. We therefore identify a dynamical analogue of mixed-order behavior rather than a new thermodynamic universality class.

The singular limit here is kinetic rather than thermodynamic: the ensemble size controls statistical precision, while the observation time controls phase-space exploration. This distinction allows comparisons with collective transitions without identifying their microscopic mechanisms.

Our analysis quantifies why the onset of exponential instability does not determine how rapidly an observable explores phase space, and establishes a general scaling criterion separating the sampled phase-space width from the diffusive exploration rate. Building on the contrasting stationary responses of focusing and dispersing boundaries, we obtain an exact local instability scale alongside a distinct, correlation-dependent transport prediction whose divergence scales as $\vert{}b\vert{}^{-3/2}$ relative to local hyperbolicity. Practically, this demonstrates that finite-time measurements near integrability can mimic effective confinement even when the asymptotic sampled width remains finite. Separately diagnosing tangent-space growth, increment correlations, and stationary spreading provides a direct framework for interpreting transport in nearly integrable Hamiltonian systems, from ray dynamics in weakly deformed optical cavities to action diffusion in area-preserving maps, and extends naturally to exact billiards and anomalous-transport regimes.

\begin{acknowledgments}
A.K.P.F. acknowledges CAPES (No.~88887.990665/2024-00), the Fulbright Program and the Fulbright Commission in Brazil -- Fulbright/CAPES Doctoral Dissertation Research Award (Process 2026--2027) for financial support. E.D.L. acknowledges support from CNPq (304398/2023-3) and FAPESP (2025/14544-0).
\end{acknowledgments}
\bibliographystyle{apsrev}
\bibliography{PRL}
\end{document}

\title{Supplemental Material for ``Continuous and Discontinuous Routes to Chaos with Diverging Transport Times''}
\author{Anne K\'etri P. da Fonseca}
\affiliation{School of Electrical Engineering and Computer Science, University of North Dakota, USA}
\affiliation{Departamento de F\'isica, Unesp -- Universidade Estadual Paulista, Av. 24A, 1515, 13506-900, Rio Claro, SP, Brazil}
\author{Marcelo de Almeida Presotto}
\affiliation{Departamento de F\'isica, Unesp -- Universidade Estadual Paulista, Av. 24A, 1515, 13506-900, Rio Claro, SP, Brazil}
\author{Diego F. M. Oliveira}
\affiliation{School of Electrical Engineering and Computer Science, University of North Dakota, USA}
\author{Edson D. Leonel}
\affiliation{Departamento de F\'isica, Unesp -- Universidade Estadual Paulista, Av. 24A, 1515, 13506-900, Rio Claro, SP, Brazil}
\maketitle
\setcounter{equation}{0}
\renewcommand{\theequation}{S\arabic{equation}}
\renewcommand{\thefigure}{S\arabic{figure}}
\renewcommand{\thesection}{S\arabic{section}}
\setcounter{secnumdepth}{1}

This supplement identifies the assumptions connecting deterministic instability to angular transport, establishes the discrete spectral origin of the singular zero-mobility limit, and derives the finite-observation and roughness-estimator results supporting the main text.

\section{Deterministic map: local versus transport time}
The parabolic end-cap shape in an unfolded cell can be written as $f(x)=4bx(x-a)/a^2$ (the opposite cap has reversed curvature and an axial translation). The tangent angle is $\arctan f'(x)\simeq4b(2x-a)/a^2$ for small slopes. Approximating the flight distance by $l$ and using the specular angular change of twice the tangent tilt gives the local reduced map
\begin{equation}
 \xi'=\left(\xi+\frac la\tan\psi\right)\bmod1,\qquad
 \psi'=\psi-\frac{8b}{a}(2\xi'-1).
 \label{eq:map}
\end{equation}
This approximation follows the established unfolded map for parabolic end caps \cite{livorati2011family}. Corrections due to finite curvature and near-grazing flights are not controlled uniformly by the small-slope approximation. The formulas below concern the smooth local branches of this map, not the entire exact billiard.

At a local fixed point with zero angular increment, $\xi^*=1/2$ and $(l/a)\tan\psi_m^*=m$, so $\psi_m^*=\arctan(ma/l)$. With $A=l/(a\cos^2\psi_m^*)$ and $B=-16b/a$, the Jacobian in coordinate order $(\xi,\psi)$ is
\begin{equation}
 J=\begin{pmatrix}1&A\\B&1+AB\end{pmatrix},\qquad \det J=1,\qquad t=\operatorname{tr}J=2+AB.
\end{equation}
Its multipliers are $\Lambda_\pm=(t\pm\sqrt{t^2-4})/2$. For $b<0$, $t>2$ and the two multipliers are real and reciprocal; for $0<b<a^2\cos^2\psi_m^*/(4l)$, $|t|<2$ and the fixed point is elliptic. At $b=a^2\cos^2\psi_m^*/(4l)$, $t=-2$ and linear ellipticity is lost. This includes the central threshold $b_c=a^2/(4l)$. None of these local statements alone establishes global mixing or the absence of islands of higher periods.

For the dispersing branch, set $h=8|b|l/(a^2\cos^2\psi_m^*)$. The positive fixed-orbit exponent is
\begin{equation}
 \gamma_m=\ln\Lambda_+=\operatorname{arcosh}(1+h)=\sqrt{2h}\,[1+O(h)],
\end{equation}
where the expansion is at fixed $m$ as $b\to0^-$. Therefore $n_{\rm inst}=1/\gamma_m\sim |b|^{-1/2}$. This asymptotic expansion is not uniform as $|m|\to\infty$ (grazing angles). If bounded, normal angular diffusion holds with $D_\psi\propto b^2$, the observed transport crossover scales as $n_x\propto |b|^{-2}$ and the ratio $n_x/n_{\rm inst}\propto |b|^{-3/2}$. This prediction differentiates the local instability timescale from the global spreading timescale even though both diverge. For the modulo-$\pi$ angular completion and $b<0$, each smooth-branch Jacobian has nonnegative entries and is entrywise greater than or equal to
\begin{equation}
 J_0=\begin{pmatrix}1&l/a\\16|b|/a&1+16|b|l/a^2\end{pmatrix}.
\end{equation}
Let $v_0$ be the strictly positive eigenvector of $J_0$ for its larger eigenvalue $\Lambda_0$, and denote the Jacobian at iteration $i$ of an orbit by $J_i^{\rm orb}$. Positivity implies, componentwise, $J_{n-1}^{\rm orb}\cdots J_0^{\rm orb}v_0\ge J_0^n v_0=\Lambda_0^n v_0$. Thus, along every orbit avoiding the singular sets, the maximal tangent-growth rate obeys $\liminf_{n\to\infty}n^{-1}\ln\|DT^n\|\ge\ln\Lambda_0=\gamma_{m=0}>0$. With modulo continuation the exceptional sets and their iterated preimages have zero area. For a finite product, the corresponding bound is $\|DT^n\|_2\ge\Lambda_0^n$ when $v_0$ is normalized in the Euclidean norm. If a maximal Lyapunov exponent exists, it is at least this bound. This makes the distinction between instability and transport explicit: for $n_x\propto |b|^{-2}$, the number of tangent e-foldings on the exploration timescale is at least of order $|b|^{-3/2}$.

\section{Correlated angular kicks and the invariant measure}
Let $u_i=2\xi_{i+1}-1$ and consider accumulated local angular kicks $S_n=(8|b|/a)\sum_{i=0}^{n-1}u_i$, prior to confinement or with a well-defined lifted angle. For a stationary fast process, let $C_b(k)=\langle(u_i-\langle u\rangle)(u_{i+k}-\langle u\rangle)\rangle$. Summing all covariances gives the finite-$n$ identity
\begin{equation}
 \frac{\operatorname{Var}S_n}{2n}=\frac{32b^2}{a^2}\left[C_b(0)+2\sum_{k=1}^{n-1}\left(1-\frac{k}{n}\right)C_b(k)\right].
 \label{eq:gk}
\end{equation}
For summable correlations and an appropriate stationary lifted increment process, the right-hand side converges to the Green--Kubo diffusivity. A preconfinement diffusion coefficient can instead be estimated from an intermediate window where the sum variance is linear in $n$. Uniform $\xi$ and negligible lagged covariances give $C_b(0)=1/3$ and $D_\psi\simeq32b^2/(3a^2)$. If the integrated correlation factor tends to a positive finite constant at small $|b|$, the exponent is two even when the prefactor is renormalized. If the factor behaves as $|b|^\delta$, then $D_\psi\sim |b|^{2+\delta}$ and the fixed-width transport exponent becomes $z=-(2+\delta)$. Anomalous diffusion or a vanishing correlation sum invalidates the simple normal-diffusion argument. The numerical collapse is evidence for an effective scaling regime.

\paragraph{Angular continuation.} Equation~(\ref{eq:map}) does not by itself keep the local coordinate $\psi$ within $[-\pi/2,\pi/2]$. Its continuation across the angular cut must be specified for a reproducible global map. If, for example, the angle is reduced modulo $\pi$ into $[-\pi/2,\pi/2)$ after the kick, the local map is a composition of two invertible shears on a product of circles (up to singular/discontinuity sets of zero area). Lebesgue measure $d\xi\,d\psi/\pi$ is then invariant. If an orbit samples this invariant measure ergodically, its temporal angular variance tends to $\pi^2/12$. Invariance of the measure alone does not establish that sampling property. Furthermore, a principal-interval angle can jump by nearly $\pi$ when a small kick crosses the cut; therefore the temporal variance of this coordinate is not a circular variance. An angular diffusivity should be calculated from the local increments or a lifted angle, not from these wrapped-coordinate jumps. A preparation away from the cut avoids this issue in the short-time limit at fixed iteration number. A circular statistic, for example $1-|n^{-1}\sum_{i=1}^n e^{2\mathrm{i}\psi_i}|$, could test sensitivity to the coordinate cut; we do not report such a test here.

For comparison, in the exact specular billiard the natural collision measure in arclength $s_c$ and signed angle $\varphi$ to the local inward normal is proportional to $\cos\varphi\,ds_c\,d\varphi$ \cite{chernov2006chaotic}. Under the corresponding normalized angular marginal, $p(\varphi)=\cos\varphi/2$, symmetry gives zero mean and
\begin{equation}
 \operatorname{Var}\varphi=\frac12\int_{-\pi/2}^{\pi/2}\varphi^2\cos\varphi\,d\varphi=\frac{\pi^2}{4}-2.
\end{equation}

This marginal concerns the angle relative to the local normal, not the axial angle $\psi$ or the end-cap return map used in the Letter.

\section{Reflected random walk: exact discrete spectrum}
Consider $x\in[-L,L]$ and a uniform step $\varepsilon Z$, $Z\in[-1,1]$. For the specular reflection convention, let $y=x+L$ and define the folded coordinate $R_L(x)=-L+2L-|((x+L)\bmod 4L)-2L|$. This triangular folding implements repeated reflection for arbitrarily long trial steps. The walk is $x_{j+1}=R_L(x_j+\varepsilon Z_j)$. This is a specific no-flux boundary implementation: rejecting a step or clipping it at a wall defines a different Markov process.

The backward transition operator on a function $g$ is
\begin{equation}
 (K_\varepsilon g)(x)=\frac12\int_{-1}^{1}g(R_L(x+\varepsilon z))\,dz.
\end{equation}
Let $\phi_m(x)=\cos[k_m(x+L)]$, where $k_m=m\pi/(2L)$, $m=0,1,\ldots$. Because $\cos(k_m y)$ is invariant under reflection at $y=0$ and $y=2L$, $\phi_m(R_L(x+\varepsilon z))=\cos[k_m(x+L+\varepsilon z)]$. Symmetry of the noise therefore gives the exact eigenvalues
\begin{equation}
 K_\varepsilon\phi_m=\rho_m\phi_m,\qquad
 \rho_m=\frac{\sin(k_m\varepsilon)}{k_m\varepsilon},\qquad \rho_0=1.
 \label{eq:spectrum}
\end{equation}
The cosine basis is complete on $[-L,L]$ and the eigenvalues are real, so $K_\varepsilon$ is self-adjoint with respect to the uniform measure. For every fixed $\varepsilon>0$, $|\rho_m|<1$ for $m\ge1$ and $\rho_m\to0$ as $m\to\infty$. The unique stationary density (among probability measures, since the cosine moments determine the measure) is $1/(2L)$. At $\varepsilon=0$, $K_0$ is the identity and every probability measure is invariant. This change in the fixed space of the evolution operator is the precise spectral meaning of the ergodicity loss in the stochastic reference.

At small $\varepsilon/L$, the eigenvalue expansion for each fixed mode is
\begin{equation}
 \rho_m=1-\frac{k_m^2\varepsilon^2}{6}+O((k_m\varepsilon)^4).
\end{equation}
Thus in the joint limit $\varepsilon/L\to0$, $n\to\infty$ with $s=\varepsilon^2 n/(6L^2)$ fixed, $\rho_m^n\to\exp(-k_m^2L^2s)$. This directly yields the Neumann heat semigroup on the interval, with $D=\varepsilon^2/6$ in iteration units. Equivalently, a Taylor expansion of the symmetric step kernel gives $(K_\varepsilon-I)g\simeq(\varepsilon^2/6)g''$ for smooth functions away from the walls. Neumann conditions follow from even reflection. In the forward equation $\partial_nP=D\partial_x^2P$, the current $-D\partial_xP$ vanishes at both endpoints, conserving probability.

The full slowest nonconstant mode for small $\varepsilon/L$ is $m=1$, with relaxation rate $-\ln\rho_1\sim\pi^2D/(4L^2)$. With $x_0=0$, however, the density remains even, and the odd modes vanish. The first active mode is $m=2$, with relaxation rate $-\ln\rho_2\sim\pi^2D/L^2$. Its inverse defines $\tau_{\rm even}\sim L^2/(\pi^2D)$. The distinction matters for prefactors but not for the perturbation exponent. The continuum density for the centered preparation is
\begin{equation}
 P(x,n)=\frac1{2L}+\frac1L\sum_{k=1}^\infty \cos\left(\frac{k\pi x}{L}\right)\exp\left(-\frac{k^2\pi^2Dn}{L^2}\right).
 \label{eq:heat}
\end{equation}
This expression is normalized, satisfies the no-flux conditions, and approaches a delta distribution at the origin as $n\to0^+$. It is valid for the diffusion approximation. In the exact discrete kernel, the exponential in Eq.~(\ref{eq:heat}) is replaced by $q_k^n$, $q_k=\rho_{2k}=\frac{\sin(k\pi\varepsilon/L)}{k\pi\varepsilon/L}$.

\section{Discrete time average and finite-observation scaling}
The time- and ensemble-averaged observable of the Letter is
\begin{equation*}
 x_\rms=\left[\frac{1}{M}\sum_{j=1}^M\frac{1}{n}\sum_{i=1}^n x_{i,j}^2\right]^{1/2}.
\end{equation*} At infinite ensemble size its square is the mean of the second moments at iterations $1,\ldots,n$. Expanding $x^2$ in the same cosine basis gives
\begin{equation}
 \frac{\mathbb{E}[x_i^2]}{L^2}=\frac13+\frac4{\pi^2}\sum_{k=1}^\infty\frac{(-1)^k}{k^2}q_k^i,
 \label{eq:moment}
\end{equation}
where $x_0=0$. Summing the geometric series yields the exact discrete, ensemble-limit formula for the specified reflected walk:
\begin{equation}
 \frac{x_\rms^2(n,\varepsilon)}{L^2}=\frac13+\frac4{\pi^2}\sum_{k=1}^\infty\frac{(-1)^k}{k^2}\frac{q_k(1-q_k^n)}{n(1-q_k)}.
 \label{eq:discrete}
\end{equation}
For $\varepsilon=0$, the fraction is replaced by its continuous limit 1, yielding zero because $\sum_{k\ge1}(-1)^k/k^2=-\pi^2/12$. For any fixed $\varepsilon>0$ the nonstationary terms vanish as $n$ grows, giving $x_\rs=L/\sqrt3$. At finite $M$, the squared estimator is unbiased for this ensemble-limit second moment, while the square-root estimator can have finite-sample bias.

The continuum scaling limit of Eq.~(\ref{eq:discrete}) produces a normalized crossover function $\mathcal{F}(s)=x_\rms/L$:
\begin{equation}
 \mathcal{F}^2(s)=\frac13+\frac4{\pi^2}\sum_{k=1}^\infty \frac{(-1)^k}{k^2}\frac{1-e^{-k^2\pi^2s}}{k^2\pi^2s},\qquad s=\frac{Dn}{L^2}.
 \label{eq:scaling}
\end{equation}
This result follows from the time average; the instantaneous second moment instead contains $e^{-k^2\pi^2s}$. For $s\ll1$, boundaries are not yet important and $\mathcal{F}^2(s)\sim s$, since the free-walk moment is $2Di$ and its discrete time average is $D(n+1)$. For $s\gg1$, the identity $\sum_{k\ge1}(-1)^k/k^4=-7\pi^4/720$ gives
\begin{equation}
 \mathcal{F}^2(s)=\frac13-\frac7{180s}+O(e^{-\pi^2s}/s).
 \label{eq:tail}
\end{equation}
Thus the time-averaged estimator has an algebraic $1/n$ saturation correction at fixed $D$, even though the instantaneous density relaxes exponentially. In particular, one should not extract a density spectral gap by fitting an exponential to this time-averaged roughness.

Matching the early and late asymptotes gives $n_x^{\rm match}\simeq L^2/(3D)=2L^2/\varepsilon^2$ in the small-step limit. A fixed-fraction criterion $x_\rms(n_f)/x_\rs=f$ for $0<f<1$ instead gives $n_f=s_f L^2/D$, where $\mathcal{F}(s_f)=f/\sqrt3$. The prefactors differ, while the divergence exponent is unchanged. 

A useful consequence concerns the apparent transition at finite observation time. At fixed fraction $f$, the perturbation at which the stationary jump becomes visible is $\varepsilon_f(n)=L\sqrt{6s_f/n}$. In particular, the jump is rounded over a window proportional to $n^{-1/2}$, even in the absence of finite-size effects. For the localized preparation,
\begin{equation}
 \lim_{\varepsilon\to0^+}\lim_{n\to\infty}x_\rms=L/\sqrt3,\qquad
 \lim_{n\to\infty}\lim_{\varepsilon\to0^+}x_\rms=0.
\end{equation}
The value at $\varepsilon=0$ is preparation dependent: for a general initial ensemble the uncentered RMS equals $\sqrt{\mathbb E[x_0^2]}$. This is why the Letter explicitly states $x_0=0$.

\section{Why the roughness and random-walk observable differ}
In the billiard the measured quantity is the ensemble average of the temporal standard deviation of each orbit, $\omega=M^{-1}\sum_j\sqrt{V_j}$, where $V_j=n^{-1}\sum_{i=1}^n\psi_{i,j}^2-(n^{-1}\sum_{i=1}^n\psi_{i,j})^2$. The random-walk quantity instead is the square root of the ensemble and time mean of $x^2$ relative to a fixed origin. Neither averaging order nor centering should be interchanged. The two observables can share exponents but do not have to share amplitudes or the entire normalized crossover function.

For illustration, in a free Brownian-like discrete process with independent, zero-mean increments of variance $2D$ starting from a fixed angle, $\operatorname{Cov}(\psi_i,\psi_j)=2D\min(i,j)$. The expected temporal variance (before boundary effects) is then
\begin{align}
 \mathbb E[V]&=\frac{2D}{n}\sum_{i=1}^n i-\frac{2D}{n^2}\sum_{i,j=1}^n\min(i,j)\notag\\
 &=D\frac{n^2-1}{3n}.
 \label{eq:var}
\end{align}
The double sum is $n(n+1)(2n+1)/6$. Hence $\sqrt{\mathbb E[V]}\sim\sqrt{Dn/3}$, whereas the fixed-origin time-averaged RMS is $\sim\sqrt{Dn}$. By Jensen's inequality $\mathbb E[\sqrt V]\le\sqrt{\mathbb E[V]}$. Brownian self-similarity gives $\mathbb E[\sqrt V]\propto\sqrt{Dn}$ in the continuum scaling regime, but with a distinct dimensionless prefactor. This accounts for the common growth exponent without identifying the two numerical observables. At $b=0$, temporal centering removes any orbit's constant initial angle. At any fixed finite iteration number and for a fixed initial angle a positive distance from the angular cut, the maximum accumulated local displacement is at most $8n|b|/a$, which tends to zero as $b\to0$. This is not uniform for initial angles approaching the cut together with $b$. This finite-time continuity holds independently of a random-phase approximation.

\section{General criterion and practical scaling diagnostics}
For a dimensionless control parameter $\lambda$, assume a single characteristic sampled width $\ell\sim\lambda^\kappa$ and normal transport with coefficient $D\sim\lambda^p$. With fixed rescaled geometry and preparation, dimensional reduction gives $W=\ell F(Dn/\ell^2)$ (asymptotically), $W\sim c_0\sqrt{Dn}$ initially, and $W_\rs\sim c_\infty\ell$. Equating asymptotes gives $n_x\sim\ell^2/D$, whence $\alpha=\kappa$, $\beta=1/2$, $z=2\kappa-p$, and $\alpha-z\beta=p/2$. A nonzero limiting accessible width, vanishing diffusivity, and a frozen unperturbed observable together suffice for a stationary jump with a divergent crossover. Width collapse instead gives continuous stationary behavior, and can coexist with slow transport when $p>2\kappa$. In the mixed-order analogy, the singularity is temporal in this fixed-domain problem; a spatial correlation-length divergence is not implied \cite{bar2014mixed,dorogovtsev2006kcore,goltsev2006kcore}.

\section{Reported numerical fits}
\begin{figure}[htb]
  \centering
  \begin{minipage}[t]{0.48\textwidth}
    \raggedright\textbf{a)}\\[2pt]
    \centering
    \includegraphics[height=5.5cm, width=\linewidth, keepaspectratio]{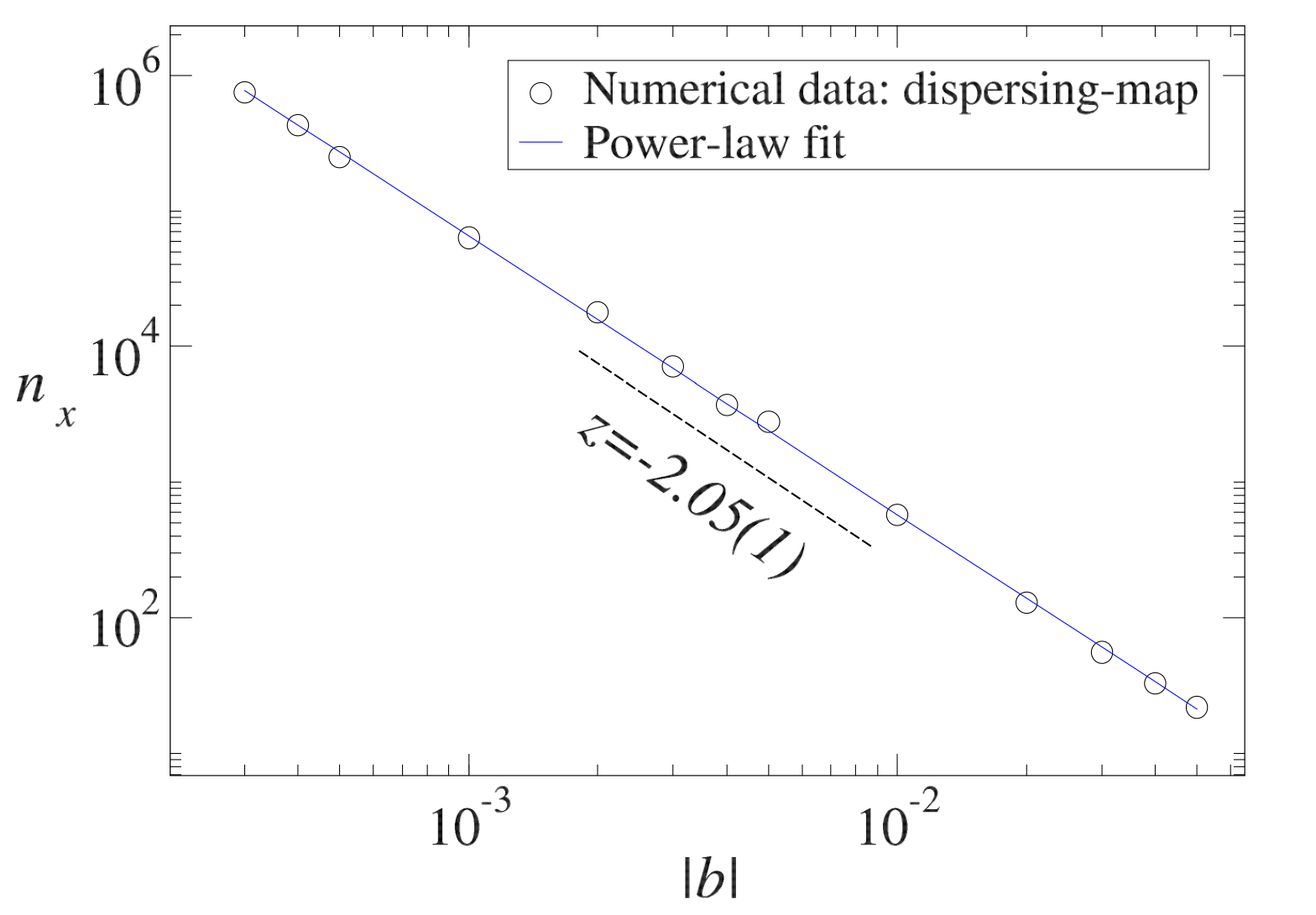}
  \end{minipage}%
  \hspace{1.2em}%
  \begin{minipage}[t]{0.48\textwidth}
    \raggedright\textbf{b)}\\[2pt]
    \centering
    \includegraphics[height=5.5cm, width=\linewidth, keepaspectratio]{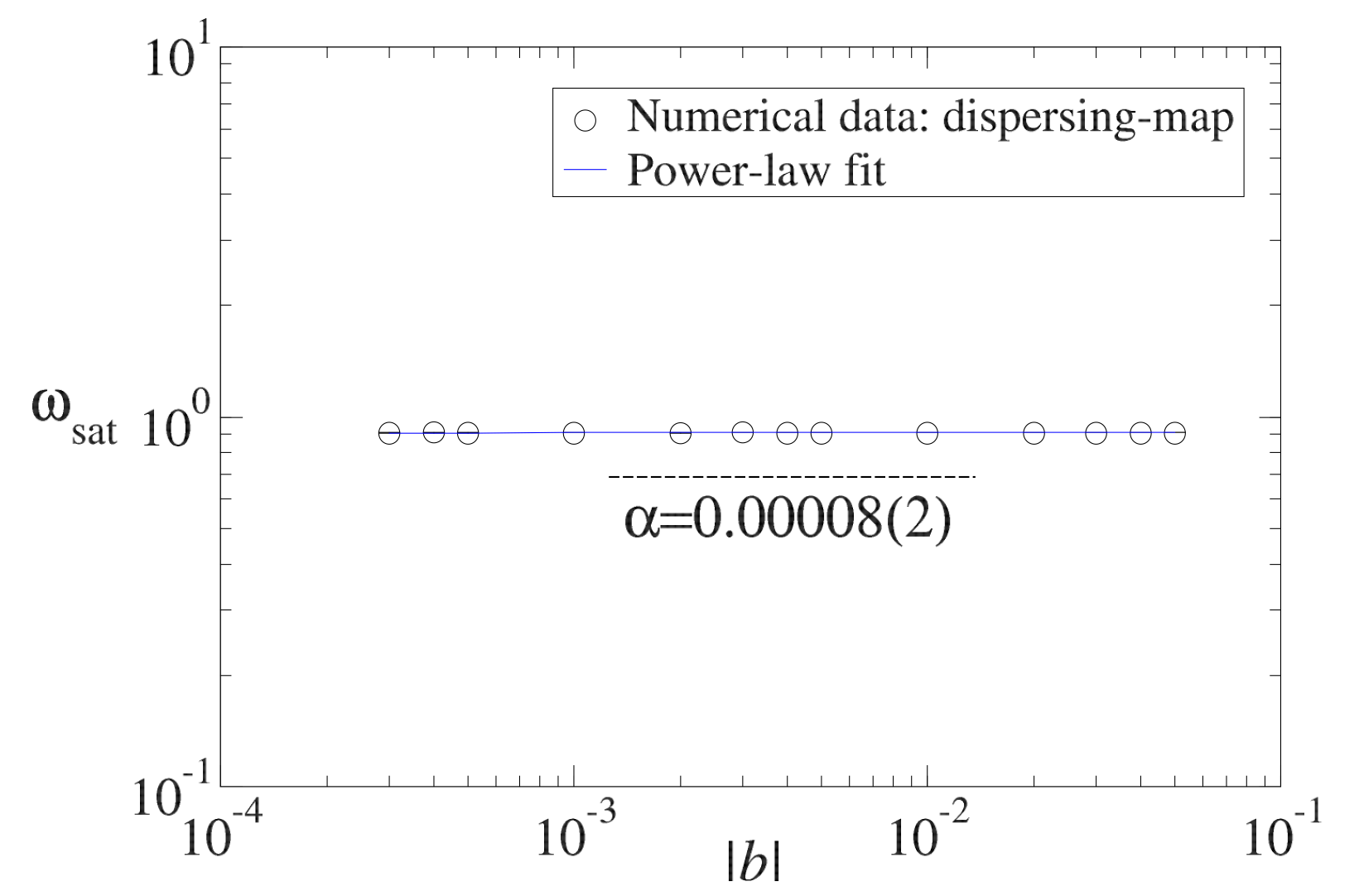}
  \end{minipage}

  \vspace{0.8em}

  \begin{minipage}[t]{0.48\textwidth}
    \raggedright\textbf{c)}\\[2pt]
    \centering
    \includegraphics[height=5.5cm, width=\linewidth, keepaspectratio]{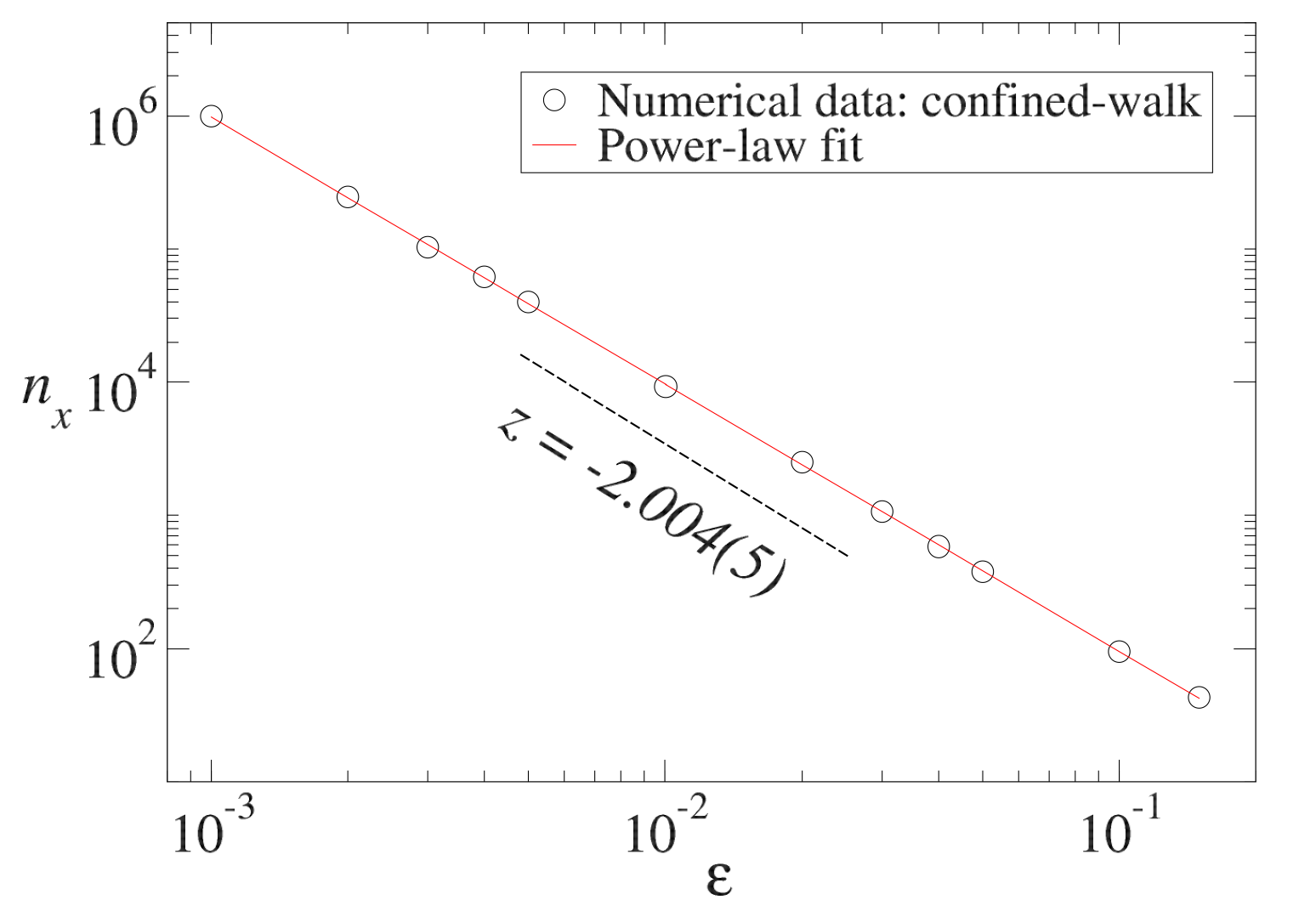}
  \end{minipage}%
  \hspace{1.2em}%
  \begin{minipage}[t]{0.48\textwidth}
    \raggedright\textbf{d)}\\[2pt]
    \centering
    \includegraphics[height=5.5cm, width=\linewidth, keepaspectratio]{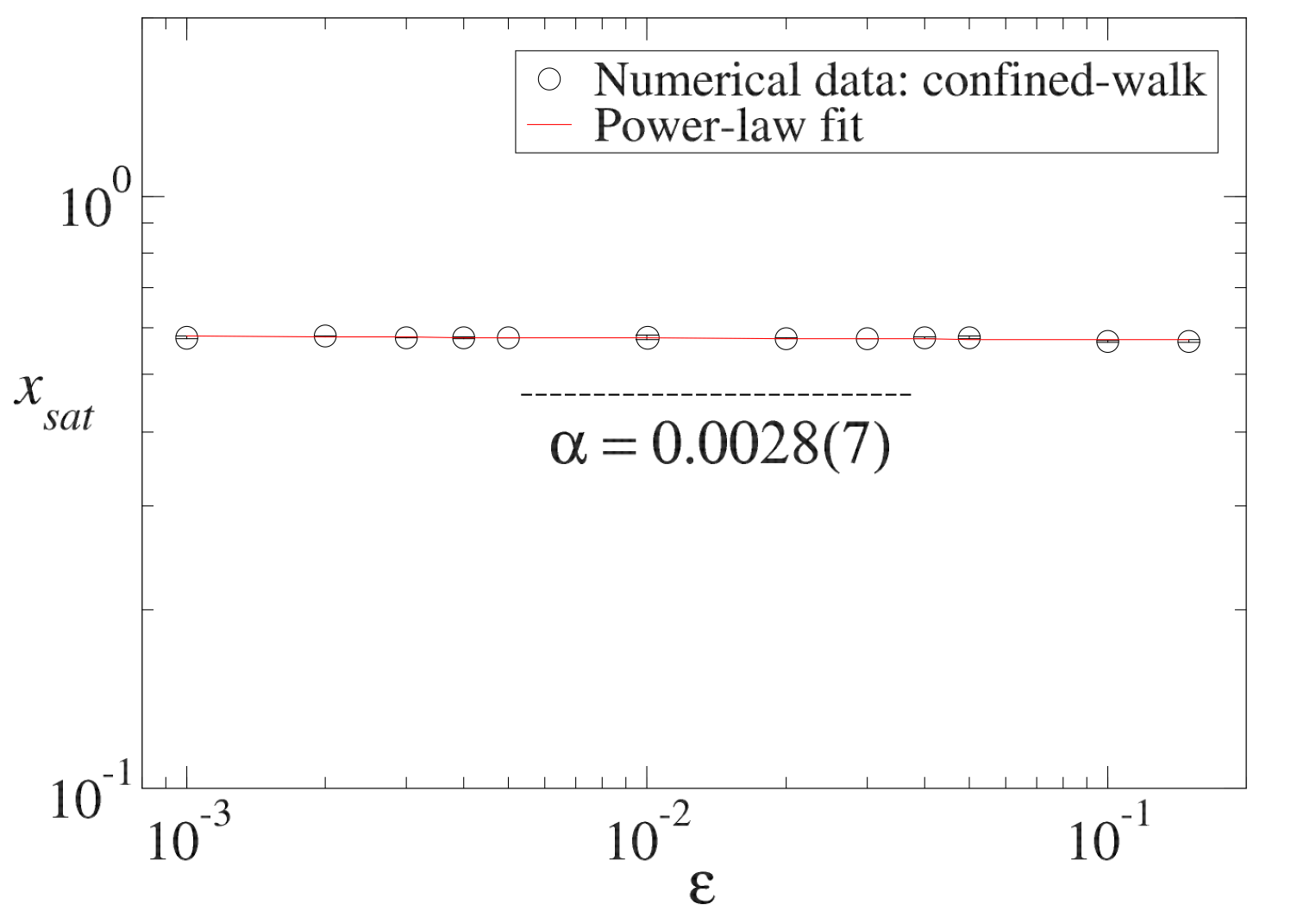}
  \end{minipage}

  \caption{Original stationary-width and crossover fits for (a), (b) the dispersing billiard map and (c), (d) the confined random walk. The printed fit values are retained without alteration.}
  \label{fig:fits}
\end{figure}

Figure~\ref{fig:fits} reproduces the original fit panels supplied for the two models, so that the numerical exponent estimates are available alongside the analytic predictions. In the dispersing map [Figs.~\ref{fig:fits}(a) and \ref{fig:fits}(b)], the growth fit in the collapse is $\beta=0.499(1)$, while the stationary-width and crossover fits are $\alpha=0.00008(2)$ and $z=-2.05(1)$. In the confined walk [Figs.~\ref{fig:fits}(c) and \ref{fig:fits}(d)], the growth fit from the original trajectory/RMS data is $\beta=0.50025(5)$, while the stationary-width and crossover fits are $\alpha=0.0028(7)$ and $z=-2.004(5)$. These values are transcribed from the original figure annotations rather than rounded or assigned revised errors.

\bibliographystyle{apsrev}
\bibliography{PRL}